\documentstyle[12pt]{article}
\begin{document}
\vspace*{6cm}
\begin{center}

ON SPECIAL CASES OF GENERAL GEOMETRY:

geometries with changing length of vectors
\end{center}
\vspace*{.5cm}

\begin{center}
{\bf{ S. S. Shahverdiyev$^*$}}
\end{center}
\begin{center}
\vspace*{1cm}

{\small   Institute of Physics, Azerbaijan National Academy of Sciences, Baku, Azerbaijan}
 \end{center}
\vspace{.5cm}


\begin{center}
{\bf{Abstract}}
\end{center}

\begin{quote}
\noindent
{\small
We find relations between quantities defining geometry and quantities defining the length of a
 curve in geometries underlying Electromagnetism and unified model of Electromagnetism and
  Gravitation. We show that the length of a vector changes along a curve in these geometries.
}
\end{quote}

\vspace{1cm}

\noindent

\noindent

\vspace{4cm}
$^*$e-mail:shervgis@yahoo.com
\newpage

\section{Introduction}
In paper \cite{sh1} a new geometry called General Geometry is
formulated  and it is shown that its the most simplest case is
geometry underlying electromagnetism. However, relation between
quantities defining geometry $F^\sigma\-_\lambda$ and the length
of a curve $A_\mu$ was assumed. Next, in paper \cite{sh2} it is
shown that geometry underlying unified model of electromagnetism
and gravitation is also a special case of General Geometry. There,
relations between quantities defining geometry
$F^\sigma\-_\lambda$,
 $\Gamma^\sigma\-_{\mu\nu}$  and the length of a curve $g_{\mu\nu}$, $A_\mu$ were also assumed.

In the present paper, it is shown that relations  assumed in
\cite{sh1} and \cite{sh2} hold to be true provided that the length
of a vector changes along a curve in both geometries.

In Riemannian geometry the length of a vector does not change and
this makes it be an underlying geometry for Gravitation. If the
length of a vector changes in Riemannian geometry then it fails to
be an underlying geometry for Gravitation. This failure has been
demonstrated
 by H. Weyl \cite{w} who investigated   Riemannian geometry with
  changing  length of a vector in an attempt to unify electromagnetism and
  gravitation (for discussion see  \cite{p}).  However, we show that the length of a vector changes
   along a curve
 in the presence of electromagnetic field  in a geometry completely
 different from Riemannian one.

In summary, Geometry of Electromagnetism \cite{sh1} with changing length
 of a vector has physical interpretation as geometry underlying Electromagnetism.
  Riemannian geometry with constant length of a vector has physical interpretation
   as geometry underlying Gravitation. Combination of Geometry of Electromagnetism
   and Riemannian geometry  with changing length of a vector is geometry
   underlying unified model of Electromagnetism and Gravitation.

In the next section we prove relations assumed in \cite{sh1} and \cite{sh2}
 and show that the length of a vector changes along a curve in
 these geometries.

\section{On Special Cases of General Geometry}

We recall that Geometry of Electromagnetism \cite{sh1} is defined
by
\begin{equation}\label{ge}
\frac{d\xi^\sigma}{du}=-F^\sigma\-_\lambda(x)\xi^\lambda.
\end{equation}
 We consider the following metric
\begin{equation}\label{ge1}
ds=\sqrt{\eta_{\mu\nu}dx^\mu dx^\nu}+\frac{q}{cm}A_\mu(x) dx^\mu, \quad \eta_{\mu\nu}=diag(1 -1 -1 -1).
\end{equation}
Accordingly, the length of a vector $V=\xi^\lambda\frac{\partial}{\partial x^\lambda}$
is
\begin{equation}\label{geN}
dl=\sqrt{\eta_{\mu\nu}\xi^\mu\xi^\nu}+\frac{q}{cm}A_\mu(x)\xi^\mu.
\end{equation}
And we assume that
\begin{equation}\label{g}
\frac{dl}{du}=\Phi_\nu(A_\lambda, F_{\mu\sigma})\xi^\nu,
\end{equation}
where $A_\mu$  are some functions of $x$ , $\Phi_\nu$ are functions of $A_{\mu}$ and $F_{\mu\nu}$, and $q$, $c$, $m$ are some parameters.
Equation (\ref{g}) means that the length of a vector changes along a curve due to $\Phi_\nu$.
Substitution of $dl$ in (\ref{g}) by (\ref{geN}) leads to equations
\begin{equation}\label{2e}
\xi^\mu\xi^\nu (F_{\mu\nu}+F_{\nu\mu})=0,\quad
\frac{q}{cm}(\partial_\mu A_\sigma x_u^\mu-A_\mu F^\mu\-_\sigma)\xi^\sigma=\Phi_\nu(A_\lambda, F_{\mu\sigma})\xi^\nu.
\end{equation}
The most general solution to the first one is any antisymmetric tensor
$$
F_{\mu\nu}=-F_{\nu\mu}.
$$
We choose $\Phi_\nu$ such that the second equation has solution\footnote{If we choose $ds=\sqrt{\eta_{\mu\nu}dx^\mu dx^\nu}$ and $\frac{dl}{du}=0$ we obtain that $F_{\mu\nu}$ is an arbitrary antisymmetric tensor and electromagnetic field $A_\mu$ has to be introduced artificially.}

$$
F_{\mu\nu}=\frac{q}{cm}(\partial_\mu A_\nu-\partial_\nu A_\mu).
$$
As it is shown in \cite{sh1},  curvature vector $R_\lambda$ is equal to
$R_\lambda=\partial^\mu F_{\mu\lambda}$. Equation $R_\lambda=0$ coincides with Maxwell equation for electromagnetic field $A_\mu$  and equation for geodesics coincides with the equation for a particle interacting with electromagnetic field $A_\mu$.
This allows us to interpret $A_\mu$ as electromagnetic field and geometry defined by (\ref{ge}) with (\ref{ge1}) and (\ref{g}) as geometry underlying electromagnetism. $q$ is identified  with charge, $m$
with mass of a particle interacting with electromagnetic field $ A_\mu$, c is the speed of the light.

If we choose $\Phi_\nu=0$ then the second equation in (\ref{2e}) reduces to
$$
\partial_\mu A_\sigma x_u^\mu-A_\mu F^\mu\-_\sigma=0.
$$
Multiplication by $A^\sigma$ gives
$$
A^\sigma\partial_\mu A_\sigma=0.
$$
This equation is a constraint for $ A_\mu$. Therefore in order to consider general functions $ A_\mu$ of $x$ we have to allow $\Phi_\nu\ne 0$.
Hence, the length of a vector must change along a curve in Geometry of Electromagnetism.

Next we consider geometry underlying unified model of electromagnetism and gravitation \cite{sh2} defined by
\begin{equation}\label{}
\frac{d\xi^\sigma}{du}=-(F^\sigma\-_\lambda(x)+\Gamma^\sigma\-_{\lambda\mu}(x)x_u^\mu)\xi^\lambda,
\end{equation}
and choose metric as
$$
ds=\sqrt{g_{\mu\nu}dx^\mu dx^\nu}+\frac{q}{cm}A_\mu(x)dx^\mu,
$$
where $g_{\mu\nu}(x)$ is a metric tensor and
 the length of a vector $V$ is
\begin{equation}\label{1}
dl=\sqrt{g_{\mu\nu}\xi^\mu\xi^\nu}+\frac{q}{cm}A_\mu\xi^\mu,
\end{equation}
and it changes as
\begin{equation}\label{2}
\frac{dl}{du}=\Phi_\nu^\prime(A_\lambda, F_{\mu\sigma})\xi^\nu.
\end{equation}
Note that in this geometry $\xi_\rho=g_{\rho\mu}\xi^\mu$.
Substitution of $dl$ in (\ref{2}) by  (\ref{1}) gives rise to
$$
\Gamma_{\nu,\sigma\lambda}+\Gamma_{\lambda,\sigma\nu}=\partial_\sigma g_{\lambda\nu},\quad
\frac{q}{cm}(\partial_\mu A_\sigma x_u^\mu-A_\mu F^\mu\-_\sigma)=\Phi_\sigma^\prime.
$$
Solutions to the first equation are
$$
2\Gamma_{\lambda,\mu\nu}=\frac{\partial{{g_{\lambda\nu}}}}{\partial
x^\mu}+\frac{\partial{{g_{\lambda\mu}}}}{\partial x^\nu}-
\frac{\partial{{g_{\mu\nu}}}}{\partial x^\lambda}.
$$
We choose $\Phi^\prime_\nu$ so that the second equation solves as
$$
F_{\mu\nu}=\frac{q}{cm}(\partial_\mu A_\nu-\partial_\nu A_\mu).
$$
According to the results obtained in \cite{sh2} we interpret $g_{\mu\nu}$ as gravitational field and $A_\mu$ as electromagnetic field.

\section{Conclusion}

In this paper we considered only two special cases of General
Geometry \cite{sh1}. Resuming, geometries  discussed in this
paper, with appropriate metrics are underlying geometries for
physical theories. The most simplest case of General Geometry
$$
\frac{d\xi^\sigma}{du}=-F^\sigma\-_\lambda(x)\xi^\lambda,
$$
with metric
$$
ds=\sqrt{\eta_{\mu\nu}dx^\mu dx^\nu}+\frac{q}{cm}A_\mu(x) dx^\mu
$$
is  geometry underlying Electromagnetism.
Next order in $x_u$, Riemannian geometry,
$$
\frac{d\xi^\lambda}{du}=-\Gamma^{\sigma}_{\lambda\nu}(x)x_u^\nu\xi^\lambda
$$
with metric
\begin{equation}\label{*}
ds=\sqrt{g_{\mu\nu}dx^\mu dx^\nu}
\end{equation}
is geometry underlying Gravitation.
Combination of two previous geometries
$$
\frac{d\xi^\sigma}{du}=-(F^\sigma\-_\lambda(x)+\Gamma^\sigma\-_{\lambda\mu}(x)x_u^\mu)\xi^\lambda
$$
with metric
$$
ds=\sqrt{g_{\mu\nu}dx^\mu dx^\nu}+\frac{q}{cm}A_\mu(x) dx^\mu
$$
is  geometry underlying unified  model of Electromagnetism and
Gravitation  \cite{sh2}.

We do not discuss the other special cases in this paper. Riemannian Geometry with metric
 $ds=\sqrt{g_{\mu\nu}dx^\mu dx^\nu}+A_\mu(x) dx^\mu$ without parameters instead of (\ref{*})
  has been considered
 in \cite{r}\footnote{I thank Prof. M. Anastasiei for informing me about \cite{r} after \cite{sh1}, \cite{sh2}
 and this paper have been posted on the Internet. It is surprising that Prof. M. Anastasiei recommended to publish \cite{bail}
 although its results contradicts those obtained in \cite{sh1} (also, see \cite{cFinsler}).} and applied to Kaluza-Klein theory.
 As we demonstrated in \cite{sh1} any attempt to geometrize electromagnetism in
geometries like Riemannian, (for example in the so called Finsler
geometry) independent of the chosen metric must fail \cite{bail}.
By choosing different metrics we do not change geometry \cite{r},
\cite{ssc}, {\cite{cFinsler}.

\section{Remarks}
Thanks to moderators of \cite{charlatan} and \cite{scienews}
almost all attempts to sabotage these serious of papers are made
available online.

\end{document}